\documentclass[11pt]{article}
\usepackage{moriond,epsfig}

\def\be{\begin{equation}}
\def\ee{\end{equation}}
\def\bea{\begin{eqnarray}}
\def\eea{\end{eqnarray}}

\def \Sleuth {{\sc Sleuth}}

\begin{document}
\vspace*{4cm}
\title{Experimental Summary Moriond QCD 2008}

\author{ A. De Roeck }

\address{CERN, 1211 Geneva 23, Switzerland\\
and University of Antwerp, Belgium}

\maketitle\abstracts{
2008 was a vintage year for the QCD Moriond meeting. Plenty of new data from
Tevatron, HERA, B-Factories and other experiments have been reported. 
Some brand new results became public just before or even during the conference. 
A few new hints for New Physics  came up in Winter 2008, 
but these await further scrutiny. This paper is the write-up 
of the experimental  summary talk given at the Moriond QCD March meeting.}

\section{Introduction}
In this paper I will discuss the progress in the different areas as reported
at Moriond QCD 2008. This year's Moriond meeting 
is very special indeed: it is the last
Moriond before the switch on of the long awaited LHC -- if all goes as planned.
Next year's Moriond meetings are very likely to contain presentations of real 
LHC data.

This year 
 is also the first Moriond meeting after the switch off of the HERA 
accelerator at DESY, Hamburg.
HERA has been a very faithful contributer to Moriond 
QCD conferences in the 
last 15 years. At Moriond 
QCD 1993, only a mere 8 months after the first timid collisions, the HERA 
experiments started to show their first QCD results. In fact, 
the first $F_2$ measurements
at low $x$ were shown at Moriond QCD 1993\cite{vallee}, and simultaneously 
at first DIS meeting in Durham\cite{albert}.
After 15 years of producing outstanding QCD results, HERA has now been 
terminated just before midnight on the 30th of June 2007. 
Many results will still be completed in the 
next few years and presented at future Moriond QCD meetings (and elsewhere).

One of the excitements at this Moriond meeting was caused by 
the possible  hints for
New Physics,
like the one from the $B_s$ decays. 
Some of the discussion of these new effects
will be developed further in the theoretical summary of Chris Quigg~\cite{quigg}.

\section{QCD}
The first topics discussed in this summary
  are the ones probably most consistent with the 
title and spirit of this conference, namely on  QCD.

\begin{figure}
\psfig{figure=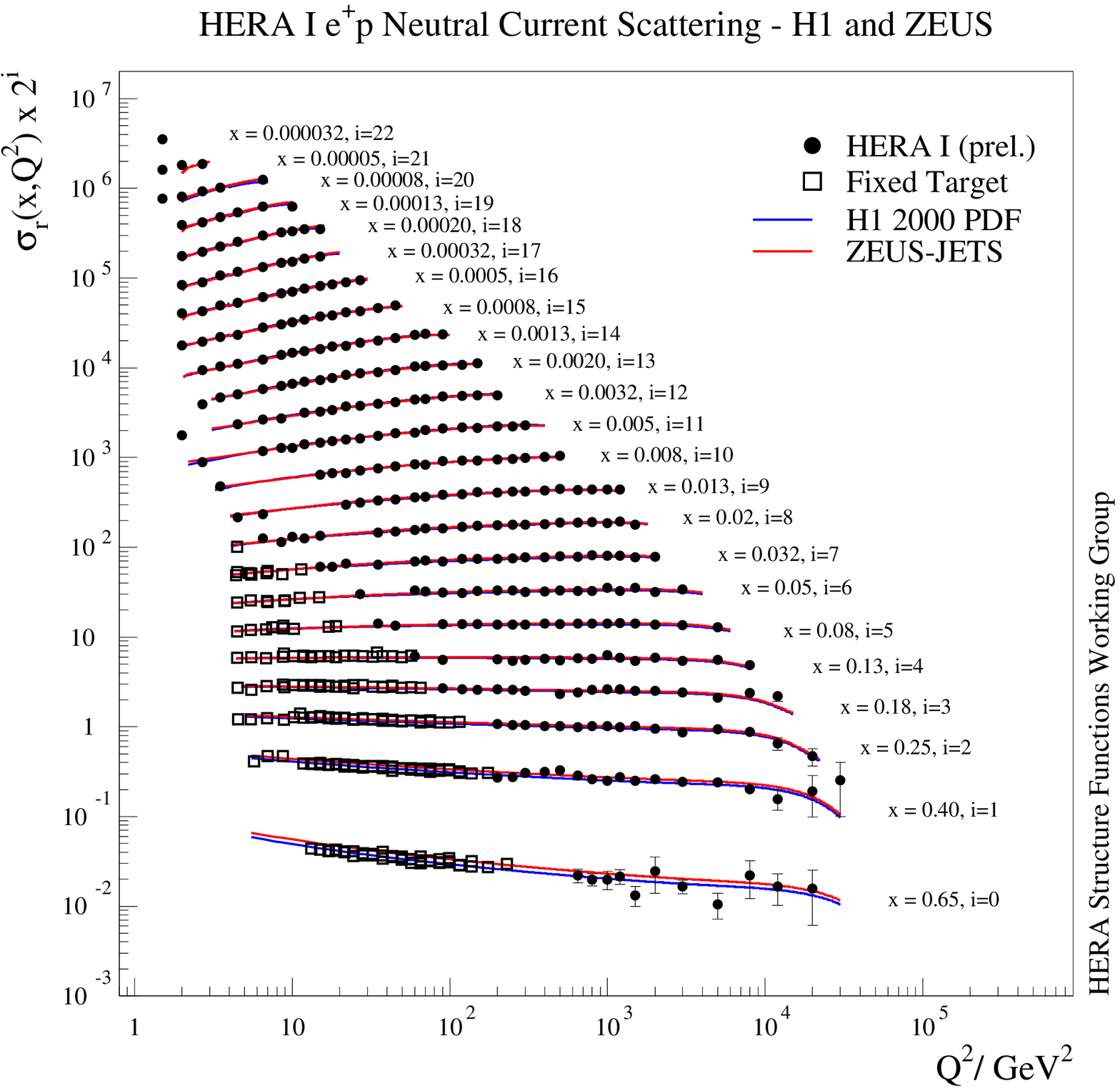,height=3.4in}
\psfig{figure=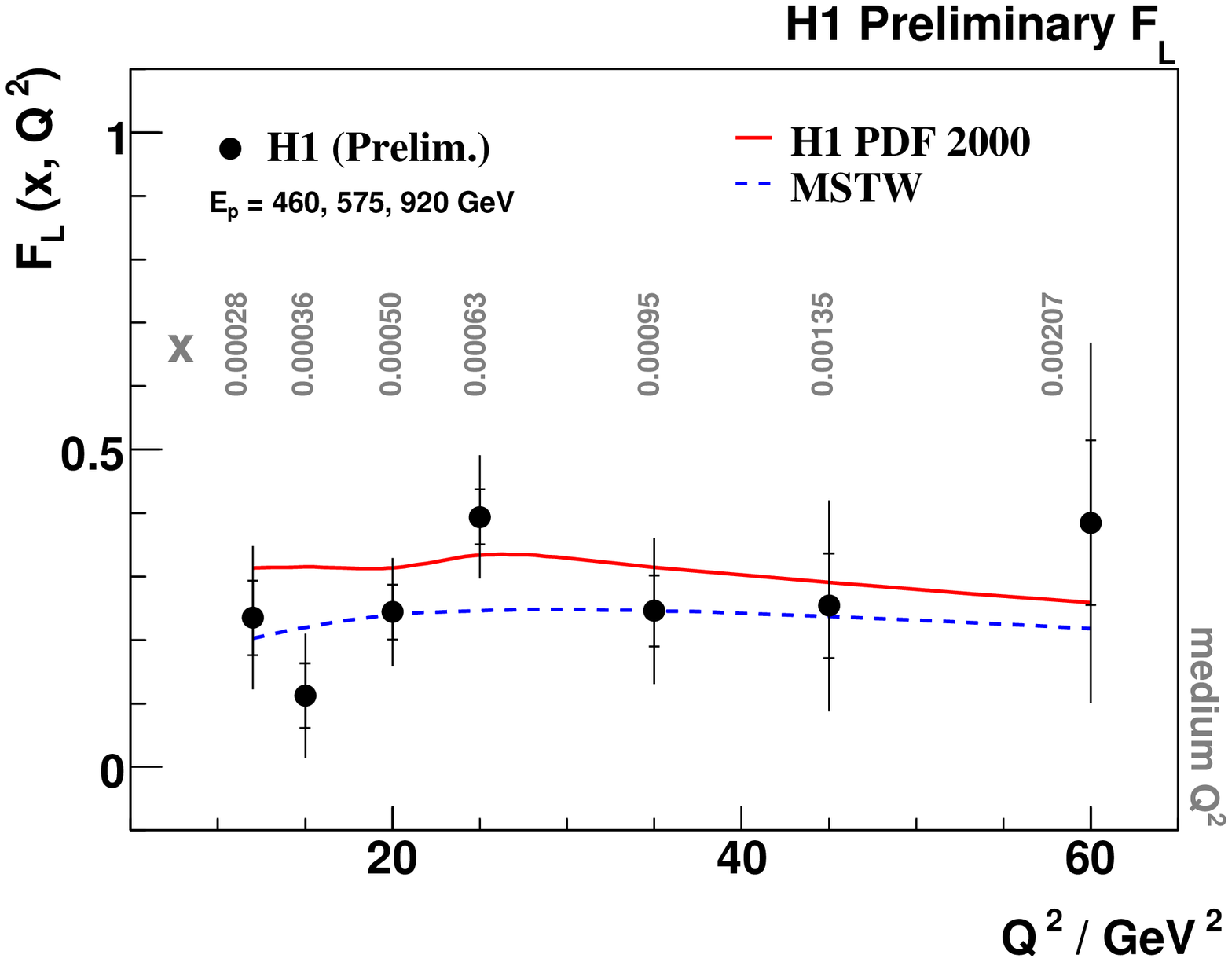,height=2.5in}
\caption{Overview of the HERA structure function $F_2$ data (left), 
and first  preliminary measurement of the longitudinal structure function
$F_L$
\label{fig:f2}}
\end{figure}

Important input for the LHC will be the understanding of the parton 
distributions of the proton. Key input to these parton distribution 
determinations are the $F_2$ structure function measurements of HERA.
The precision of the HERA $F_2$ data is now 1-3\% in the bulk region, but still 
statistics limited for 
the largest $x$ and $Q^2$ values, see Fig.~\ref{fig:f2}.
At this 
Moriond meeting a direct measurement of the second structure function $F_L$  
 was released for the first time~\cite{nikiforov}, see 
Fig.~\ref{fig:f2}.
During the last 3 months of operation HERA ran at reduced proton energy
(at two different energies, namely 460 and 575 GeV) and  combining these 
 data with the data at
920 GeV allows to extract $F_L$. 
Another  recent development is the combined structure function data set
from the two HERA experiments\cite{gorzo}, 
ie combining the ZEUS and H1 $F_2$ data and using clever 
techniques to cross calibrate the systematics. These combined
measurements have reached a truly fantastic precision, and
during the HERA-LHC workshop\cite{heralhc} 
on May 2008 the power of these combined structure
functions $F_2$ in PDF extractions was shown, 
reducing the parton uncertainties by a factor of two or so in a large region.
Getting the best PDFs for the LHC is one of the ongoing challenges and has
recently condensed in a forum to stimulate that work, called 
PDF4LHC~\cite{PDF4LHC}.

Jets are another set of  classical QCD measurements, and 
several new jet measurements were shown~\cite{jiminez} at this meeting; 
an example are the  mini-jet measurements for jets with $p_T > 3$ GeV. 
These measurements are likely 
to be important for helping to understand 
the dynamics of the underlying event data at the 
LHC.
Inclusive jet measurements are now also being included in PDF analyses. 
This 
particularly helps to additionally constrain the gluon at high and medium $x$.
It also allows to extract precise values at $\alpha_s$ as discussed 
in the theory summary talk.

\begin{figure}[ht]
\begin{center}
\epsfig{file=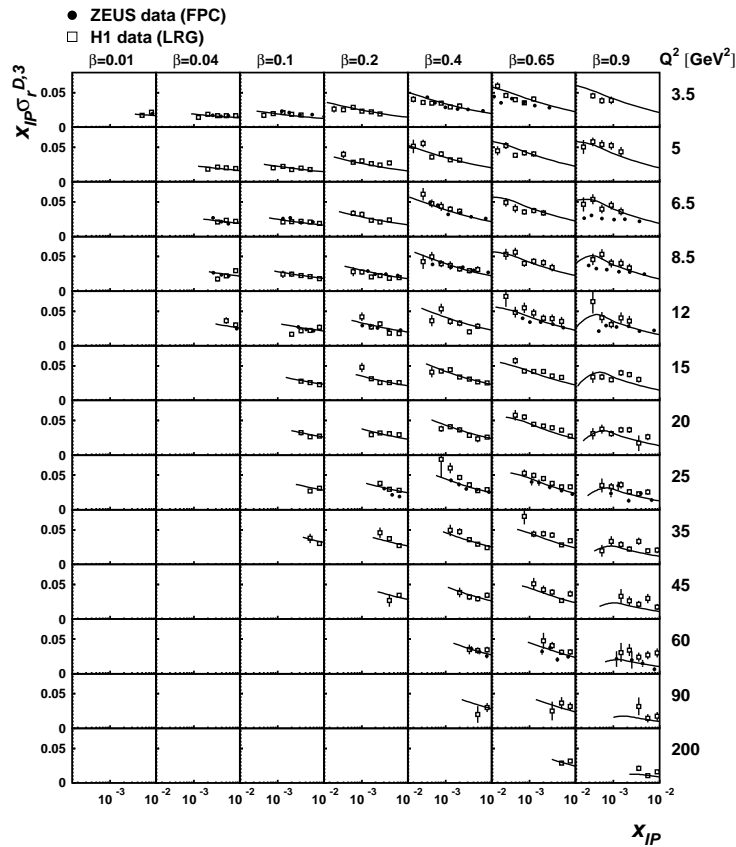,width=0.6\textwidth} 
\caption{Measurement of the diffractive structure function by H1 and ZEUS
and the theoretical prediction from\protect \cite{marget} }
\label{fig:diff}
\end{center}
\end{figure}

\begin{figure}
  \begin{center}
    \psfig{figure=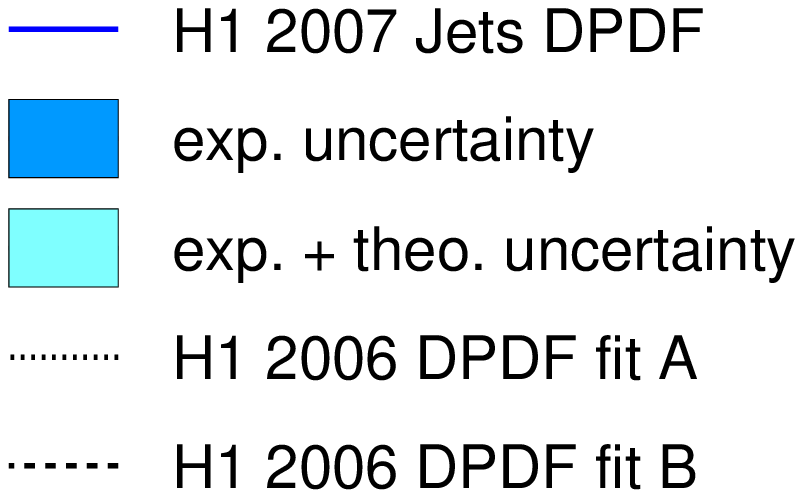,height=25mm}
    \psfig{figure=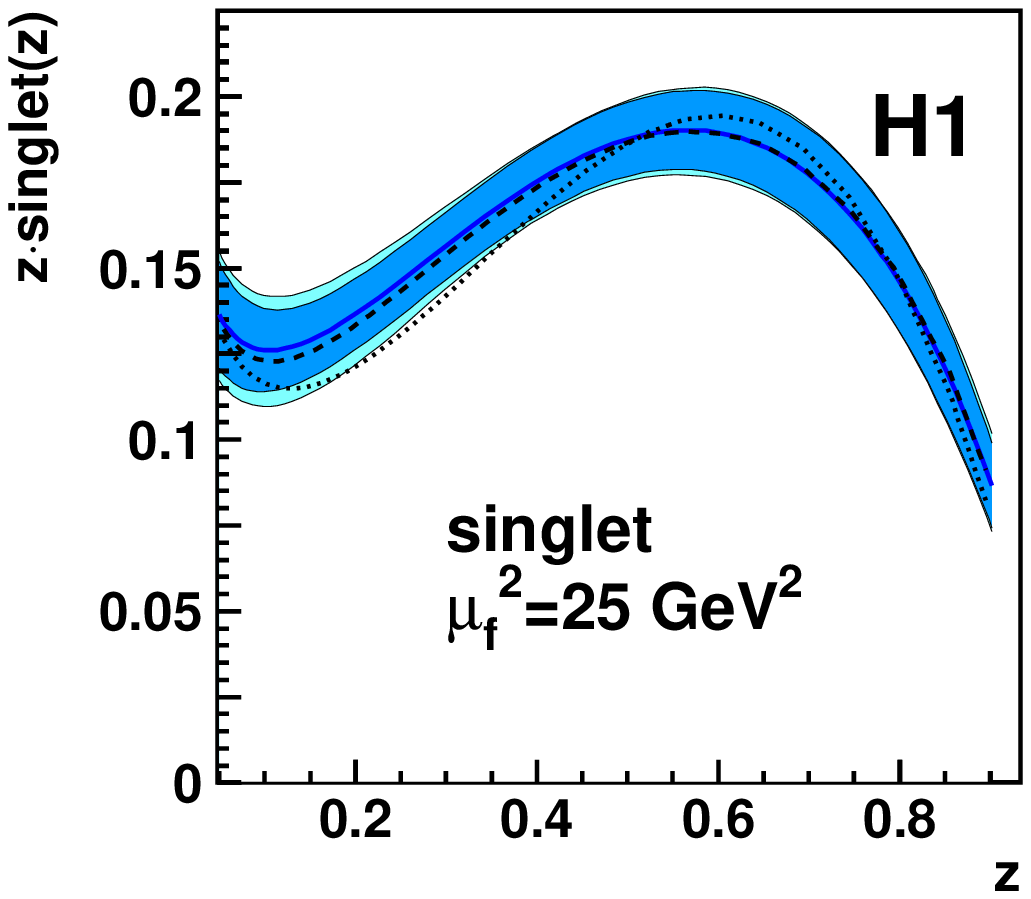,height=42mm}
    \psfig{figure=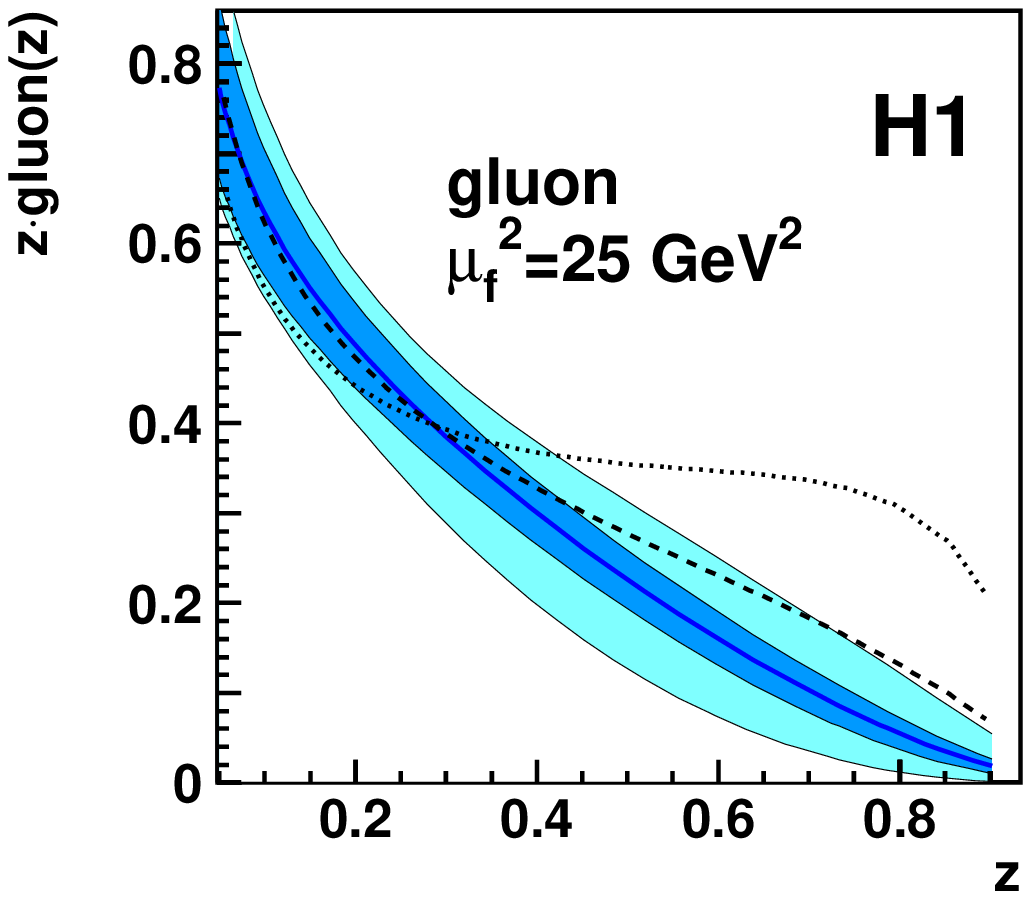,height=42mm}
  \end{center}
  \caption{The diffractive quark density (left) and the diffractive
    gluon density (right) versus $z$, the momentum fraction of the 
    parton,
    for the squared factorisation scale
    $\mu ^2 _f = 25$ GeV$^2$. }
  \label{fig:dif2}
\end{figure}

A third strong leg of HERA QCD measurements
is provided by  the diffractive data. The diffractive 
structure function $F_2^D$ is   measured precisely~\cite{lim},
 as shown in Fig.~\ref{fig:diff}.
Several different methods are used by the experiments for
 extracting $F_2^D$ .
Some notable differences between the H1 and ZEUS data, 
eg at high $\beta$ and $Q^2$ of about 5-10 GeV$^2$,
are present. A phenomenological analysis carried out in~\cite{marget}
 offers a parameter free prediction of the diffractive cross sections, 
and --amusingly-- does seem to referee between the "off bins" in the data 
sets. However there is no over-all winner, in some bins (high $\beta$) the 
H1 data are preferred, but at $\beta \sim 0.4$ the model
 seems to prefer the 
ZEUS data.
The diffractive structure function data is also used to extract parton
distributions of diffractive exchange, as shown in Fig.~\ref{fig:dif2}.

\begin{figure}
  \begin{center}
    \psfig{figure=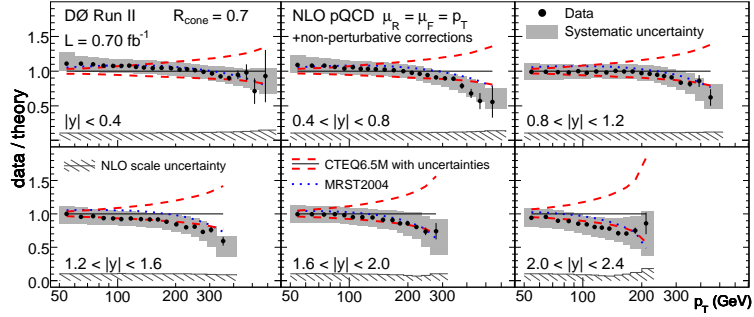,height=42mm}
  \end{center}
  \caption{Measured data divided by theory for the inclusive jet 
  cross section as function of $p_T$ in several $y$ bins.}
  \label{d0jets}
\end{figure}

The Tevatron has been delivering impressive data on jet measurements
in the last years. CDF
and D0 showed recent precision jet measurements for jet $p_T$ values up to 
600 GeV~\cite{brown}. Some of these recent measurements are shown in 
Fig.\ref{d0jets}.
Again these measurement will help to constrain 
the gluon at high $x$ in PDF studies, and are now being 
incorporated in the global fits.
 These measurements add value for the PDFs studies on
top of the HERA jet measurements, since they are generally at a higher 
scale (several hundreds of GeV$^2$) than the HERA ones.

A particularly important measurement for LHC studies is the reported 
result on exclusive di-jet production in events with rapidity gaps
(aka diffractive events). CDF discussed  the ratio of the distribution of the 
invariant mass of the di-jets over the invariant mass
of all objects observed in the central detector. 
The amount of events at values above 0.6 of this ratio
can only be understood if exclusive dijet events, i.e. events with only 
two jets in the central detector, are added to signals in the Monte Carlo.
Hence this demonstrates (together with other channels such as exclusive
di-photon production) that exclusive processes exist at high energies.
Moreover the observed amount of exclusive events is close to the prediction
of the Durham group~\cite{khoze}.
The main interest in this channel for the LHC
is the exclusive production of the Higgs boson, as will be discussed later.

Results on jet+photon data have been discussed at the meeting. These 
measurements have a long history of "avoiding agreement with theory" and 
this still seems to be the case for the latest measurements~\cite{sonnenschein}.
The recent D0 data show  a $\sim 20\% $ lower cross section than the theoretical
expectation for photon $p_T$ 
values larger than 100 GeV. This may well be bad 
news for the LHC, where one counts on this process for PDF and other QCD 
studies. 
Originally this process was expected to be more reliable at high $p_T$ values.
However, CDF data seems to be more in accord with the theory in this region.
Remarkably the photon+b-quark jet seems to agree already with the LO
calculations. Just luck? More precise data will show.

New heavy flavour data, in the QCD context, were presented by both HERA
and the Tevatron. The latest results on $F^b_2$ shows that the $b$-content of
the proton is about 1\%. This measurement is important for the 
determination
the Higgs production cross sections at the LHC, especially in extensions
of the Standard Model (SM). 

A few years ago at the Tevatron there
were discrepancies between the measured $b$-quark spectra
and the theoretical predictions. 
Meanwhile these differences have been ironed out but the question remained 
whether the modifications applied would also work at other energies (say LHC)
or other processes (say $ep$ scattering). The HERA  $p_T$ spectra 
of the $b$-quark jets~\cite{meyer}
 shed light on this issue: indeed the calculations work 
reasonably well for HERA, "from the first shot". Some discrepancies at low 
$p_T$ values are  observed. More precise data will be the referee 
here, but it is likely that some additional theoretical work may be 
needed for the low $p_T$ region.

New Tevatron data resolve the outstanding
puzzle on the di-muon cross section: the 
CDF run-I measurement  
was significantly higher than the expectation but the new 
CDF Run-II  measurements are  in agreement with the
NLO calculation. The $J/\psi$ polarization data are found not to be described 
by NRQCD calculations~\cite{pursely} (In the discussion it was claimed that 
the Durham calculations can however describe these data\cite{khoze2}).
In all: how well do we really understand heavy quark production at the
Tevatron and HERA? Can we safely extrapolate to the LHC?
There clearly are some areas where more insight is needed.

\begin{figure}[t!]
\begin{center}
\includegraphics[width=0.75\textwidth]{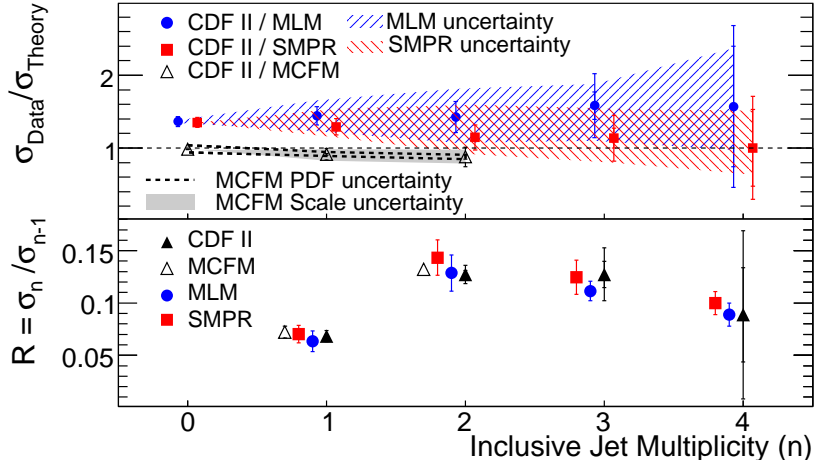}
\end{center}
\caption{
The ratio of data to theory for the total cross
sections 
as a function of the jet multiplicity $n$.
Bottom: $\sigma_n/\sigma_{n-1}$ for data, MLM, SMPR and {\sc MCFM}
calculations.
Inner (outer) error bars denote the statistical (total) uncertainties
on the measured cross sections.}
\label{fig:wzjets}
\end{figure}

Of particular relevance for the LHC is the understanding of vector boson
production (+jets) at the Tevatron. Processes with vector bosons will 
constitute an important background  to  many searches for new physics,
notably for supersymmetry.
CDF reported a first observation of the $ZZ$ process at the Tevatron: 
 a signal is seen with $4.4\sigma$ significance, based on 3 clear events 
with basically zero background~\cite{lipeles}.
Remarkable are the results on W+jets, Z+jets, W+$c$jets 
and Z+$b$jets measurements.
The W+jet measurements are now made for up to 4 jets, as shown in 
Fig.~\ref{fig:wzjets}. The measured 
cross sections are in excellent agreement with the NLO MCFM predictions, 
essentially straight out of the box, for up to two jets. 
This is excellent news for the LHC 
and leaves us to hope that we will understand the W+jets background at the 
LHC fast. However there is a caveat: the MCFM predications are 
at the parton level and 
the data are corrected to the hadron level only, so the agreement may not
be as impressive as it looks at first sight~\cite{neu}, but still it
is still very close.

Time for a few electroweak measurements from HERA. During Run -II the 
electrons and positrons beams 
of HERA could be polarized to roughly 60\%. This can 
be used to search for right-handed currents or make measurements on the 
axial and vector couplings of the $u$ and $d$ quarks.
It can also be used to set limits on the quark radius; and the limit is now
$R_q < 0.74\cdot 10^{-18}$m at 95\% CL.

Beam polarization in QCD can further 
be used to make measurements of the proton spin structure. 
The study of the
longitudinal spin decomposition of the proton is still an active field,
 and new results from the proton-proton collider RHIC, with polarized 
beams, were reported at this meeting. 
Using combined measurements from
STAR and PHENIX,  for jets and $\pi^0$s, 
the $A_{LL}$ asymmetry constrains the  polarized gluon 
$\Delta G$ to be $-0.8 < \Delta G < 0.2$ with 90\% CL in the range of 
$0.02< x < 0.3$~\cite{sakuma}. 
This is a good constraint for the various models and predictions.

One of the mysterious observations in QCD are the large transverse single 
spin asymmetries. These have been 
established at low center of mass (cms) energy collisions 
over 10 years ago,  and recently got 
confirmed at the RHIC collider at the highest 
cms energies\cite{seidl} for polarised collisions: $\sqrt{s}$ = 200 GeV . 
The asymmetries of the process $pp^{\uparrow} \rightarrow \pi X$ 
are studied at different Feynman-$x$ values
with a single transversely polarized proton beam. The asymmetries increase with
$x_F$ and reach values as large as 
 0.1. The results are compatible with zero at small
and negative $x_F$. It was noted~\cite{bunce} 
that there is a stringent prediction from QCD that can be checked namely 
Sivers(DIS) = $-$Sivers(DY). Hence, one should go out and confirm this 
prediction!

\begin{figure}[htb]
\center{\psfig{figure=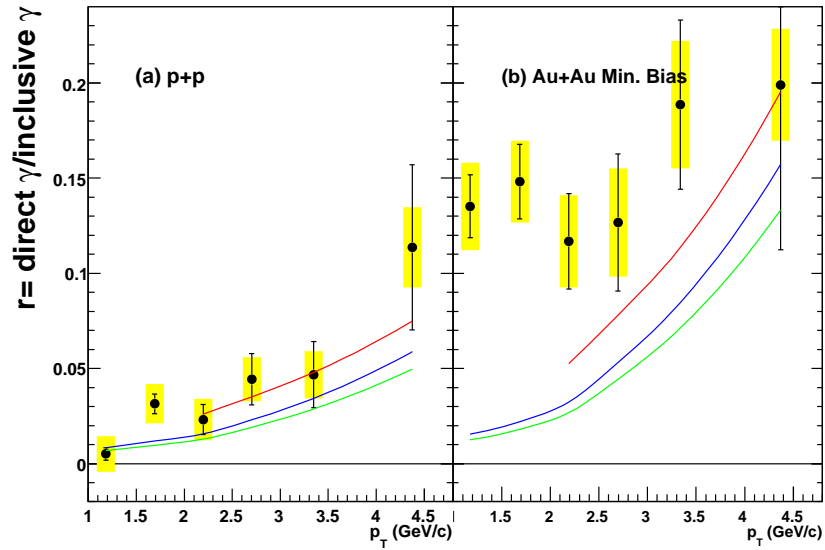,height=3.20in}}
 \caption{The obtained fractions of the virtual direct photon component as a    
function of $p_{T}$ in p+p (left) and Au+Au (right) collisions.
\label{fig:photon}}
\end{figure}

\begin{figure}[htbp]
\begin{center}
  \includegraphics[height=8cm,width=10cm]{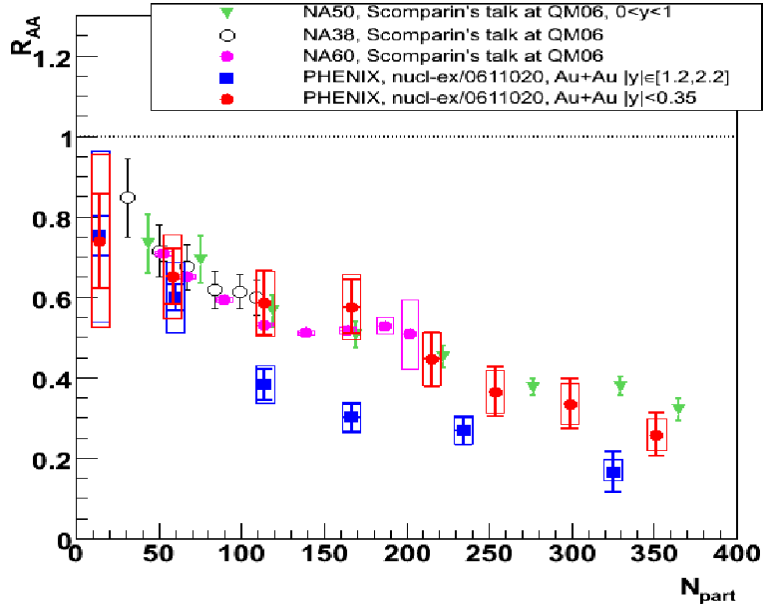}
\end{center}
\vspace{-2ex}
\caption{ J/$\psi$ R$_{AA}$ vs. N$_{part}$ at SPS compared to
  RHIC.
\label{fig:jpsi}}
\vspace{-2ex}
\end{figure}

\section{Heavy-Ion Collisions}
The RHIC heavy ion collider was conceived in order 
to establish a new state of matter 
in heavy ion collisions (aka the quark gluon plasma). Recent years have 
provided 
a wealth of data and measurements in e.g. gold-gold collisions.
Results on thermal dilepton pairs were discussed 
in~\cite{yamaguchi,floris}. As shown
in Fig.\ref{fig:photon} the $pp$ data
 seem to be consistent with NLO calculations, while $AuAu$ data are
 systematically 
above the predictions. 
For the resonance measurements, the $\rho$ gets wider and 
there is an excess in the region $M_{\mu\mu}$ which is not due to charm 
production. Inspecting the $T_{eff}$ shows a rise at small invariant mass,
consistent with the radial flow of a hadronic source, while the drop at 
large invariant mass indicates a partonic source.

Jet quenching has been observed since a number of years and is further 
studied in detail. Taking one jet as the trigger jet (near side) one can
eg. study the cone angle of the second jet (away side). The result 
disfavours Cerenkov radiation as the main effect of the quenching.
It is still unclear what the dynamics of the "ridge" at the near side is. 
It behaves as the inclusive part but more correlation studies are 
ongoing~\cite{bielcikova}.

Three particle correlations are studied with jet variables~\cite{Holzmann}.
Presently the correlations are found to be consistent with conical emission
but the presence of other jet topologies cannot be ruled out yet.
Finally, hard probes are being studied, such as heavy flavours and the 
colour charge effect~\cite{mioduszewski}. 
Correlations are studied e.g. looking at the 
nuclear enhancement of $N_{part}$ for a certain photon $E_T$ trigger,
where a suppression is seen for high $N_{part}$~\cite{atomassa}. 
A puzzling part in the $J/\psi$ suppression data at RHIC is the 
PHENIX data that show that the more central part of the production is 
LESS suppressed than the more forward part, by almost a factor of 2!
The data are shown in Fig.~\ref{fig:jpsi}.
This is a priori counter intuitive and brings up the fear that perhaps the 
$J/\psi$ may be NOT a good probe for the study of the new state of matter.
An approximate formulae for $dP_{\pi}/dx_E$ was discussed 
in~\cite{tannenbaum}

Clearly a lot of progress was made over the last years
in understanding the state of matter 
that is created in high dense systems. This new state  seems to act as 
a perfect liquid. But the data show that we cannot yet be fully satisfied
with our understanding, and more detailed and sophisticated correlation studies
are expected to shed more light on the dynamics. In other words, we are 
well en route, but not quite there yet.

\section{Heavy Flavours}

The harvest of heavy flavour physics
 from BaBar, Belle, CLEO, Tevatron is very rich.
The B-factories collected now about 1.3 ab$^{-1}$ together. 
Samples of in total of about $10^{12} B_{u,d}$ decays, $10^6 B_s$ decays and 
a few times $10^7 \psi(2s)$ decays\cite{barsuk} are available now. Heavy 
flavours are a way to probe new 
physics through appearance of the new particles in the loops. To discover 
new physics this way luminosity is crucial.
I will be relatively brief in this section since much of that is picked up
in~\cite{quigg}.

\begin{figure}[htbp]
\begin{center}
\includegraphics[height=8cm,width=11.cm]{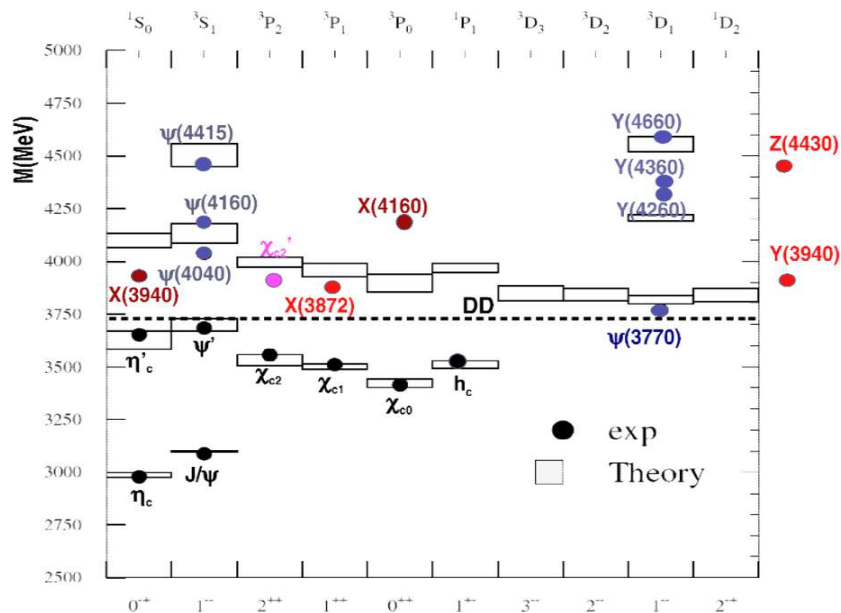}
\end{center}
\caption{ Spectrum of new states studied at Belle an BaBar (from 
\protect \cite{mussa})
\label{fig:spectrum}}
\vspace{-2ex}
\end{figure}

\begin{figure}[h]
\centering
\epsfig{figure=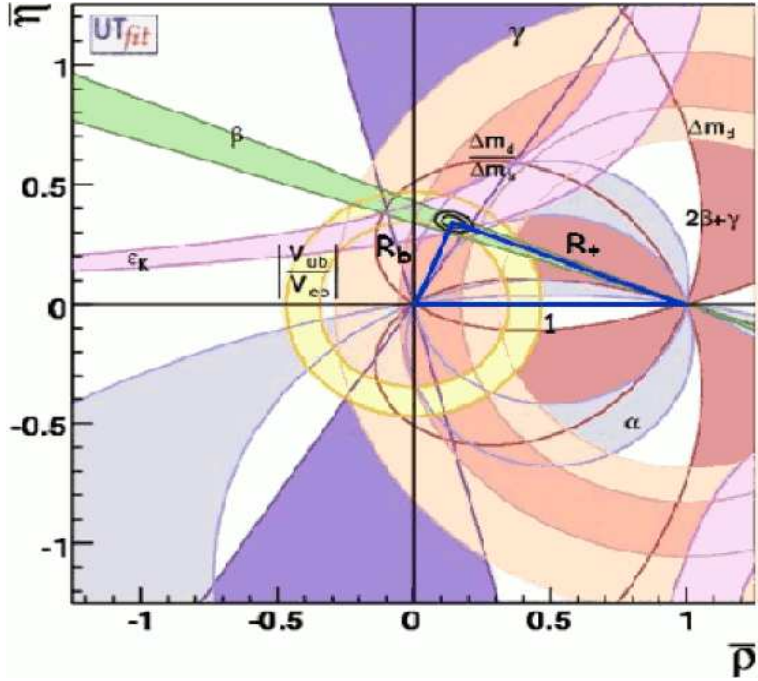,width=10.cm,height=9cm}
\caption{Update of the unitarity triangle constraints$^2$}
\label{fig:triangle}
\end{figure}

Belle and BaBar reported on the new charmonia that have been 
observed~\cite{mussa,grauges,chen}, several of which are candidates for new
 states.
An overview picture is given in Fig.~\ref{fig:spectrum}.
The $\tau$ and charm decay studies have been reported~\cite{lange}. 
There is evidence for  new
$\Xi_c$ states with masses of 3055 and 3122 MeV respectively. The earlier
discovered states at 2980 and 3077 MeV have been confirmed. 
New quarkonium results from BaBar include a measurement of the B meson mass
difference: $m(B^0)-m(B^+) = 0.33\pm 0.05\pm 0.03 $ MeV which is compatible 
with the world average but the error is a 
factor 4 reduced w.r.t. previous measurements. The significance
 for a non zero mass difference is now larger than $5\sigma$~\cite{arnaud}.
The hadronic B decays from BaBar and Belle  showed evidence for
direct CP violation from a Dalitz plot analysis of $B^{\pm} \rightarrow
K\pi\pi$ at the level of $3\sigma$\cite{lombardo}.
An update of the unitarity triangle
 is shown in Fig.~\ref{fig:triangle}, which includes 
improvements due to results  from the B-factories and the 
Tevatron~\cite{barsuk}. The precision on the angles is now roughly 
$\alpha \sim 8^0, \beta \sim 1^0$ and $\gamma \sim 13^0$.

The Tevatron showed recent measurements on masses and lifetimes of hadrons 
containing $b$-quarks. $\Xi_b$ mesons are now well established and CDF 
made the measurement of the mass to be $5792.9\pm 2.5(stat.)\pm 1.7(sys.)$
MeV. The $B_s$ lifetime measured in $B_s\rightarrow J/\psi \phi$ is now 
1.52 ps with an error of a few \% as measured in CDF and D0~\cite{Niamuddin}.

Charm mixing was reported exactly a year ago for the first time
from BaBar and Belle, and has now also been observed at the Tevatron
in CDF, with a 3.8$\sigma$ significance disfavoring the no-mixing scenario.
The elongated error ellipses are large for the different experiments and do 
not have the same central points in the so called $x',y'$ space, but are
claimed to be all compatible.

About two years ago, reported at the Moriond meetings for the first time,
the first measurement of the $B_s$ oscillation was shown to the world by 
D0. Meanwhile both CDF and D0 have shown 
more evidence and improved the results.
The experiments now report~\cite{Weber}
\begin{itemize}
\item  CDF: $\Delta m_s = (17.77\pm0.10\pm0.07) ps^{-1}$
\item  D0: $\Delta m_s = (18.53\pm0.93\pm0.30) ps^{-1}$
\end{itemize}
Also measurements on  $|V_{td}|/|V_{ts}|$ where reported  
which are now dominated by theoretical uncertainties. The personal world 
average calculated by the rapporteur is 
$\delta m_s= (17.78\pm 0.12) ps^{-1}$ and $|V_{td}|/|V_{ts}|=
0.2059\pm 0.0007 (exp)^{+0.0081}_{-0.0060} (theor)$.

New results on rare B and charmed meson decays were reported in~\cite{kuhr}. In particular 
the decay $B_s\rightarrow \mu\mu $ generates considerable interest. The limits 
of D0 and CDF are now respectively 7.5 and 4.7  $ \times 10^{-8}$,  derived 
with 2 fb$^{-1}$ of data. The SM expected value is 3.4 $\times 10^{-9}$, and 
if e.g. SUSY exists one should see the decay well before that, hence the 
Tevatron experiments are closing in on it!
For the decay $B\rightarrow\mu\mu $ the present Tevatron and 
B-factory limits are still more than 2 orders of magnitude away from the 
SM limit.
Finally, a first 
direct CP violation measurement of hadronic charmless $b$ baryon decays
was reported by CDF.

In connection with the interpretation of the g-2 experiment at 
BNL (for further discussion see ~\cite{quigg}), it is very important to measure 
the $e^+e^-$ cross section at low $\sqrt{s}$. Such measurements can be made 
but they are very tricky. It was shown that measurements are under 
way~\cite{grauges} in BaBar, using radiative events,
 but it may take some time before all final states 
in the 1-2 GeV energy region are analysed. The hope is that when it all
comes together one could have the total cross section determined with 
a precision of about 1\%.

CLEO showed an analysis of the 2007 data in the cms energy range of 7-10 GeV.
The data below 8 GeV showed a large discrepancy between the different 
experiments (Mark-I, Crystal Ball and MD1)~\cite{skwarnicki}.
 The very precise CLEO data 
referees that region and shows that the Mark-I data may well 
suffer from a systematic effect since the R value is about 20-25\% larger
compared to the new precise measurements.

Staying with the CLEO data, we hit the first serious hint for New Physics 
at this Moriond meeting. CLEO~\cite{stone} 
has made a precise measurement of the
leptonic decay constant for $D_s$ mesons: $f_{Ds}$
equal to $274\pm 10 \pm 5$ MeV. This constant can be calculated on the 
lattice and in fact a precision determination exists, which shows that there
is a 3.8$\sigma$ discrepancy between the calculation and the 
data~\cite{zwicky,kronfeld}. Can one take this discrepancy seriously?
A discussion on the theory part is given in~\cite{quigg}. If indeed this 
is a real effect then a  natural explanation 
could be given by leptoquarks in the mass range of 700-800 GeV. 
Other possible scenarios include new Wprimes or 
charged Higgses.

In the week before the conference, the UTFIT collaboration
reported on first evidence 
for new physics in the $b$ to $s$ transitions\cite{utfit}: an analysis 
of  $B_s\rightarrow J/\psi$ decays measured 
at the Tevatron experiments has found
a disagreement between the observed mixing amplitude $\phi_s$ and the SM
prediction at the 3.7 $\sigma$ level. This lead to a discussion at the 
conference both in and outside the sessions. All agree that there is indeed 
tension in the present data. Not everybody agrees on the claimed 
significance. At this point it is perhaps more a hint than evidence, and 
the jury is still out for the final verdict. 
CKMfitters await more input  on the data
from CDF/D0, and both experiments themselves
are engaged in making their own fits.
So watch that space!

\begin{figure}    
\centering \psfig{figure=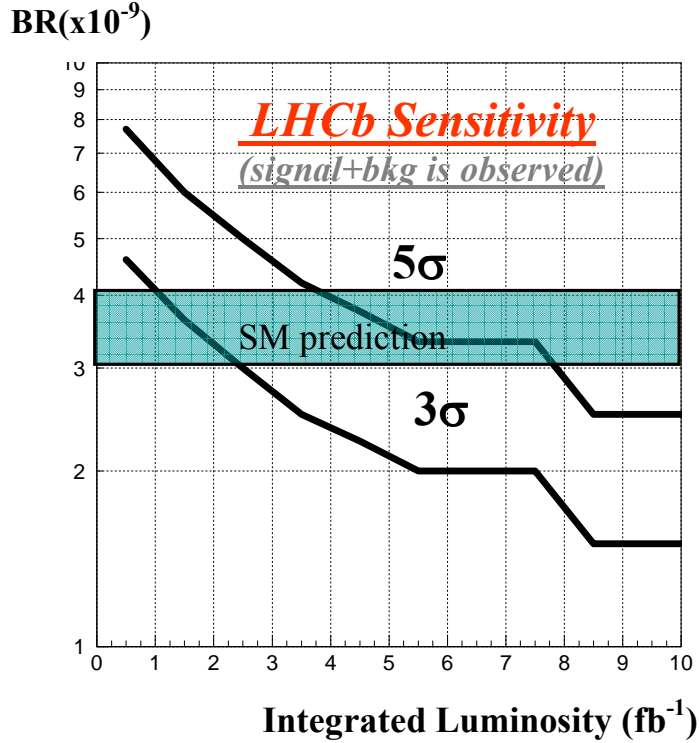,scale=1.6}
\caption{Observation ($3\sigma$) and Discovery ($5\sigma$) limits
for $B_s\to\mu^+\mu^-$ as a function of integrated luminosity in 
case where both signal and background are observed.
\label{fig:lhcb}}
\end{figure}

An important next player in this field will be 
LHCb~\cite{hicheur}. LHCb can measure 
$\phi_s$ with a precision of about 0.02 with 2 fb$^{-1}$
and can clear up the status of the 
 discrepancy. 
Note that LHCb can also measure the 
$B_s\rightarrow \mu\mu$ of the level of the SM with 0.5 fb$^{-1}$, hence 
it should be a referee on both issues already within the first 
1-2 years of physics data (ie 2009-2010). The expectations for LHCb
are shown in Fig.~\ref{fig:lhcb}.

Finally a third hint of new physics was discussed, the so called 
$\Delta A_{k\pi}$ puzzle, and discussed in the theoretical 
summary of this meeting\cite{quigg}.

\section{Top Quark Physics}
The Tevatron is still the only place in the world where the top quark 
is produced in the laboratory. Not for much longer, however!
The 3.5 fb$^{-1}$ delivered luminosity/experiment at the Tevatron
 is good for 22K produced top pairs.
The analyses presently use between 0.9 an 2.3 fb$^{-1}$.

\begin{figure}[h]
\centering
\epsfig{figure=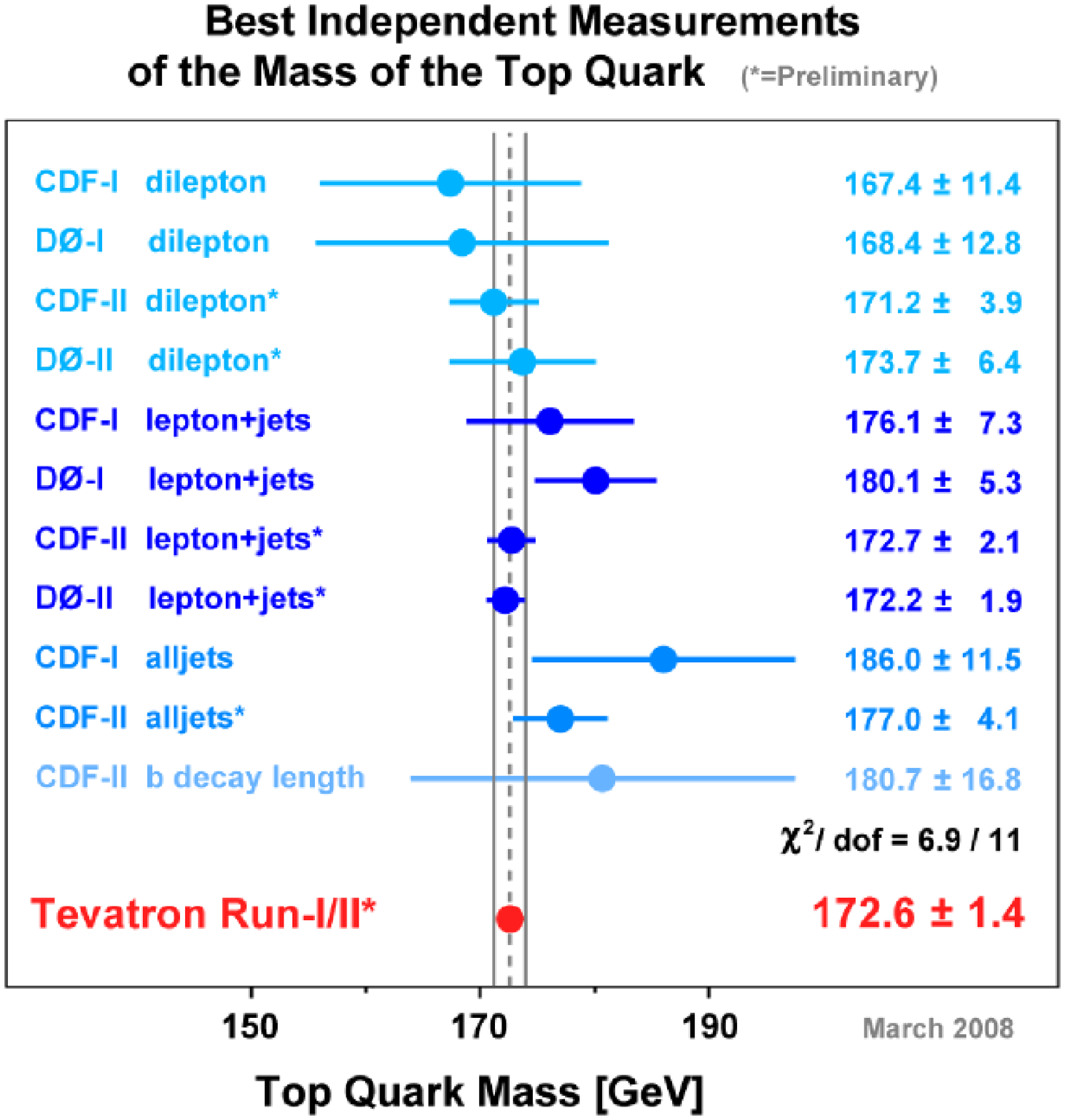,height=5.5cm,width=7.cm}
\epsfig{figure=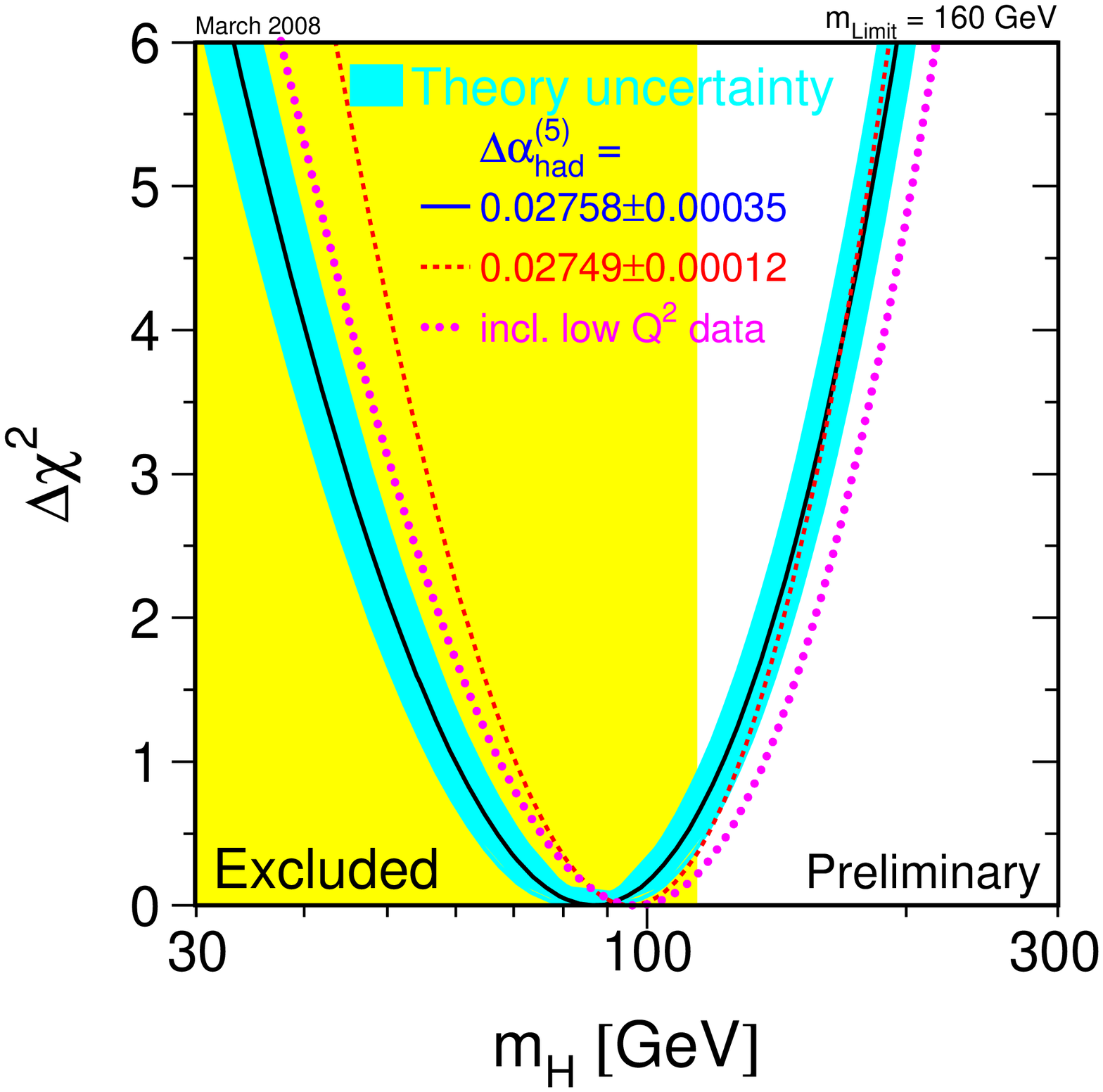,height=5.5cm,width=7.cm}
\caption{(left) Summary of the top quark mass measurements at the Tevatron;
(right) summary of the fit of the electroweak data with the new 
top mass\protect\cite{grunwald}. }
\label{fig:top}
\end{figure}

The top analyses at the Tevatron are truly impressive! At this conference 
a new value of the top mass was presented. The 
reported value is ~\cite{heintz}:
$m_{top}= 172.6 \pm 1.4$ GeV, hence $\delta m_{top}/m_{top}$ = 0.8\%.
It looks like the Tevatron experiments will succeed to reach a 
$\delta m_{top}$ of 1 GeV by 2009 or so. Hence the LHC will have a hard time
competing with these results. But the large statistics at the LHC (factor 
100 more per fb$^{-1}$) will pay off at the end by allowing for more 
stringent selections and leaving room for more ingenious methods, yet to
be developed. A summary of the top mass measurements at the Tevatron, 
and the new summary of the fit of the electroweak data~\cite{grunwald}
are given in Fig.~\ref{fig:top}.

A check was presented of the precision on the mass that could be reached 
from the cross section of the top quark production. Presently that looks 
like a factor of 5 worse\cite{thery} than achieved with the methods above.
The new top mass measurement 
was included in the electroweak fits~\cite{grunwald} and the 
following fit results were obtained (Table~\ref{table1}).

\begin{table}[t]
\begin{center}
\begin{tabular}{lcl}
-$\delta \alpha_{had} $ &  = & $0.02767\pm0.00034$ \\
-$ \alpha_s $&=& $0.1185\pm0.0027$\\
-$ M_Z $ & = & $91.1874\pm 0.0021 $ GeV \\
-$m_{top}$ & = & $172.8\pm 1.4$ GeV \\
-$m_{Higgs}$ & = & $ 87+36-27 $ GeV \\
\end{tabular}
\end{center}
\caption{Current values of the parameters of the fit of the electroweak data
with the new top mass}
\label{table1}
\end{table}
 
Further top quark properties have been  studied~\cite{Eusebi} and reported.
The decay branching ratio of top to $Wb$ is larger 
than 79\% at 95\% CL, and the 
branching ratio to the decay $B(t\rightarrow Zq)$ is less than 3.7\% at 
95\% CL. The lower mass limit on a 4th generation t' is now 284 GeV
at 95\% CL. Top charge is consistent with the Standard Model and exotic
models are excluded with 87\% CL. Helicity measurements are consistent
with SM expectations but have still 30-50\% uncertainties and leave 
room for a surprise. In any case, as far as we can see, the top behaves 
pretty much as expected "for a top quark".

Top pair production comes dominantly from $q\overline{q}$ production
at the Tevatron, with only a fraction of about $0.07 +0.15-0.07$ coming from 
gluon-gluon processes. 
This is well known to be quite different at the
LHC. The total cross section from a combination of all the channels
is quoted by CDF to be $7.3\pm0.5\pm0.6\pm0.4$ pb where the errors are
statistical, systematical and lumi respectively. This is consistent with 
the theory prediction, which is between 6 and 7.5 pb for a top mass of 175 GeV.
A search for charged Higgs production in top decays $(t\rightarrow H^+b
\rightarrow csb)$ with a charged Higgs mass of 80 GeV shows that 
$B(t\rightarrow H^+b) < 0.35 $ at 95\% CL~\cite{cortiana}.

Last year the first evidence of single top production was reported. It has 
turned out to be very difficult to extract the signal in the 
environment of high SM background processes, but the Tevatron 
experiments have succeeded to do so. Many special statistical techniques
have been deployed to find the signal (matrix elements, decision trees,
Bayesian NN, likelihood functions etc.) 
and the interconsistency of the 
results of the different methods has generated confidence in the initial 
result. By now more data has been included in these studies (eg CDF updated 
with 2.2 pb$^{-1}$) and the analysis techniques got better tuned.
The single top signal seems well established now in both CDF and D0 with 
a significance larger that 3 $\sigma$~\cite{jian}.
The cross section is about 2 pb measured in CDF and about 4.7 pb in D0.
Due to the large uncertainties (30-50\%) the measurements are both still 
consistent with the theoretically expected value of about 3 pb.

\section{Higgs Searches}
\begin{figure}[h]
\centering
\epsfig{figure=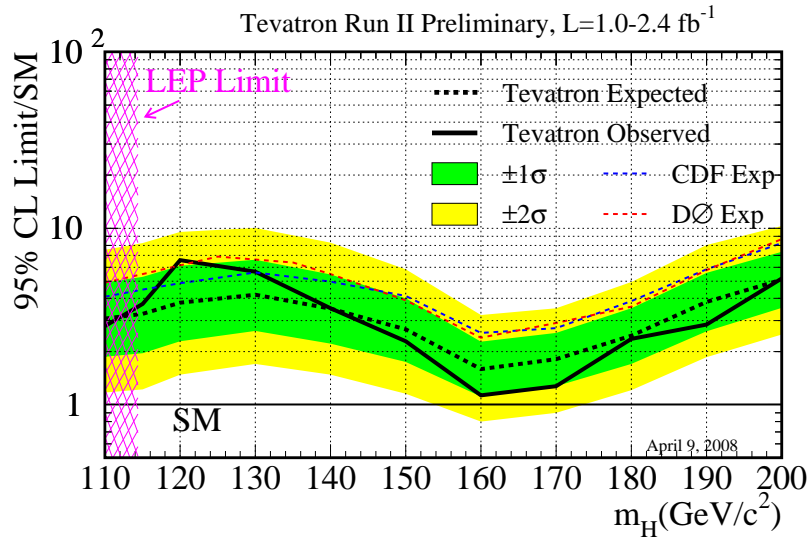,width=12.cm}
\caption{The combined exclusion plot for the Higgs from the Tevatron data.}
\label{fig:higgs}
\end{figure}

The whole world is waiting for the turn on of the LHC, to start the 
ultimate and decisive hunt for the so far elusive Higgs particle. This
"God particle", coined like that by L. Lederman because it created 
diversity in what would otherwise be a dull Universe, is often thought
of as  the 
last missing piece of the Standard Model. It is responsible for the 
electroweak symmetry breaking in the SM, telling us why e.g. Z and W bosons
are so heavy. The whole world is waiting? 
Not quite: in a region in Batavia, IL, USA, 
there is brave "gaulois" resistance to the upcoming reign of the LHC over 
this region, and 
all possible efforts are made to get to the Higgs
 before the LHC turns into 
routine physics operation.
Many channels are studied at the Tevatron~\cite{mommsen}, e.g.:
$W/H\rightarrow e\nu/\mu\nu +bb, Z/H\rightarrow e\nu/\mu\nu +bb,
Z/H\rightarrow \nu\nu +bb, W/H\rightarrow \gamma\gamma,
H\rightarrow \tau \tau, H\rightarrow WW\rightarrow l\nu l\nu$
and new results were reported at this meeting 
on most channels. A new Tevatron combination was
made for QCD Moriond 2008, which is presented in Fig.~\ref{fig:higgs}
for a 95\% CL exclusion limit compared to the SM expectation.
The remarkable thing to note is that, perhaps due to a lucky 
downward fluctuation,
the observed limit at 160 GeV starts to get close the the SM expectation, i.e. 
if this continues the Tevatron could exclude that region 
--or discover the Higgs!-- before the 
search at the LHC starts in earnest. The region around 160 GeV is the 
one where the LHC could make a discovery with a few 
100 pb$^{-1}$, i.e. very early on. Theorists at the conference made a plea to look also at 
signals down to 100 GeV masses or lower, despite the LEP limit of 114 GeV,
and give the combination plot also with signal and error.

The Higgs to $\tau$ decays was looked at with special attention, generated since 
last year's upward fluctuation in the visible mass spectrum of the two 
$\tau$'s in CDF, which could be consistent with an $M_A$ of 160 GeV.
Later that year D0 reported no excess in that channel (in fact if anything,
 a deficit), and adding new statistics also now in CDF the spectrum is 
"back to normal"\cite{wright}. Updates on the $3b$ channel, which shows a 
slight deviation as well, are coming soon.


Various other channels such as $H\rightarrow \gamma\gamma, 
H^{++}H^{--} \rightarrow
\mu^+\mu^+\mu^-\mu^-, H\rightarrow aa\rightarrow \gamma\gamma$
have been looked at, but no smoking gun was found\cite{mulhearn}.
Also a fourth generation seems to be excluded for a Higgs in the mass range
of 130 to 195 GeV at 95\% CL.

\begin{figure}[htb]
\begin{center}
\includegraphics[width=4.0in]{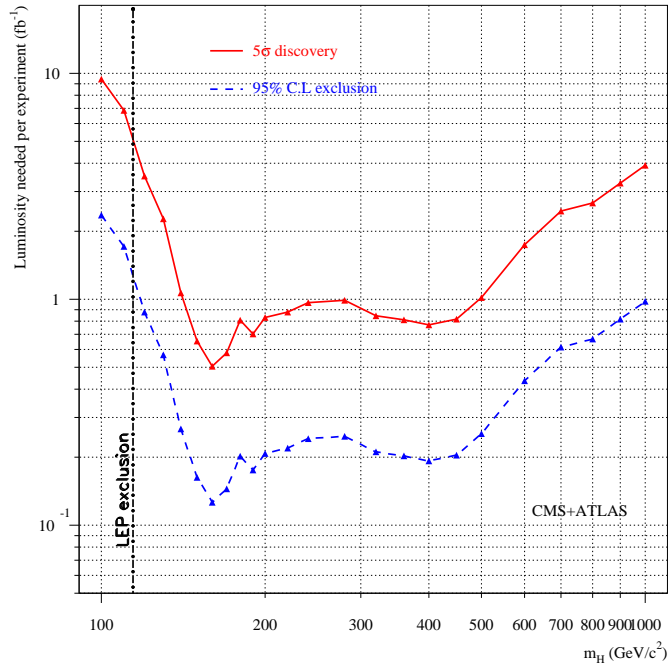}
\end{center}
\caption{The prospects  
for discovering a Standard Model Higgs boson in initial LHC
running, as a function of its mass, combining the capabilities of ATLAS
and CMS. From\protect\cite{jj}.}
\label{fig:higgslhc}
\end{figure}

Bring in the LHC~\cite{Vickey}! Clearly, the ATLAS and CMS experiments have 
been tailored for the discovery of the Higgs. The expected discovery plot for 
the combination of the two experiments is shown in 
Fig.~\ref{fig:higgslhc}, and
shows that about 1 fb$^{-1}$ of well understood CMS plus ATLAS combined 
data can be sufficient 
to discover the Higgs except when the mass is 130 GeV or below, or above
500 GeV, which 
will need more data. A review of the LHC capabilities for Higgs discovery
is reported in~\cite{polesello}.
In all, the prospects for the LHC are excellent to answer the by now
over 40 years old question: does the Higgs particle (and field) exist or not?
But there are always killjoys: in \cite{bij} it is argued that it may well be
 that Higgs particle may not be detectable at the LHC at all because 
it will be too broad a state ... More on that 
is discussed in~\cite{quigg}.

\section{Searches for New Physics}
The Tevatron continues to push for searches for new particles and 
new phenomena~\cite{strogolas,eads}. So far these searches are negative
(otherwise the content of this summary would have been quite different).
Table~\ref{table2} gives the present approximate limits on the masses 
for SUSY particles.

\begin{table}
\begin{center}
\begin{tabular}{ll}
Chargino mass (mSUGRA)& $\sim 140-150$  GeV\\
NL neutralino mass (mSUGRA)& $\sim 140-150$ GeV \\
Chargino mass (GMSB)  & $\sim 230$ GeV \\
LSP Neutralino mass (GMSB) & $\sim 125$ GeV\\
Chargino mass (mSUGRA) RPV & $\sim 200$ GeV \\
Neutralino mass (mSUGRA) RP V& $\sim 100$ GeV \\
Squark mass & $\sim 400$ GeV \\
Gluino mass & $\sim 300$ GeV \\
Light stop or RPV stop mass & $\sim 150$ GeV \\
Stop as CHAMP & $\sim 250 $ GeV\\
\end{tabular}
\end{center}
\caption{ Approximate limits on the masses
for SUSY particles from the Tevatron searches}
\label{table2}
\end{table}

\begin{figure}
\begin{minipage}{0.5\textwidth}
  \centering
  \psfig{figure=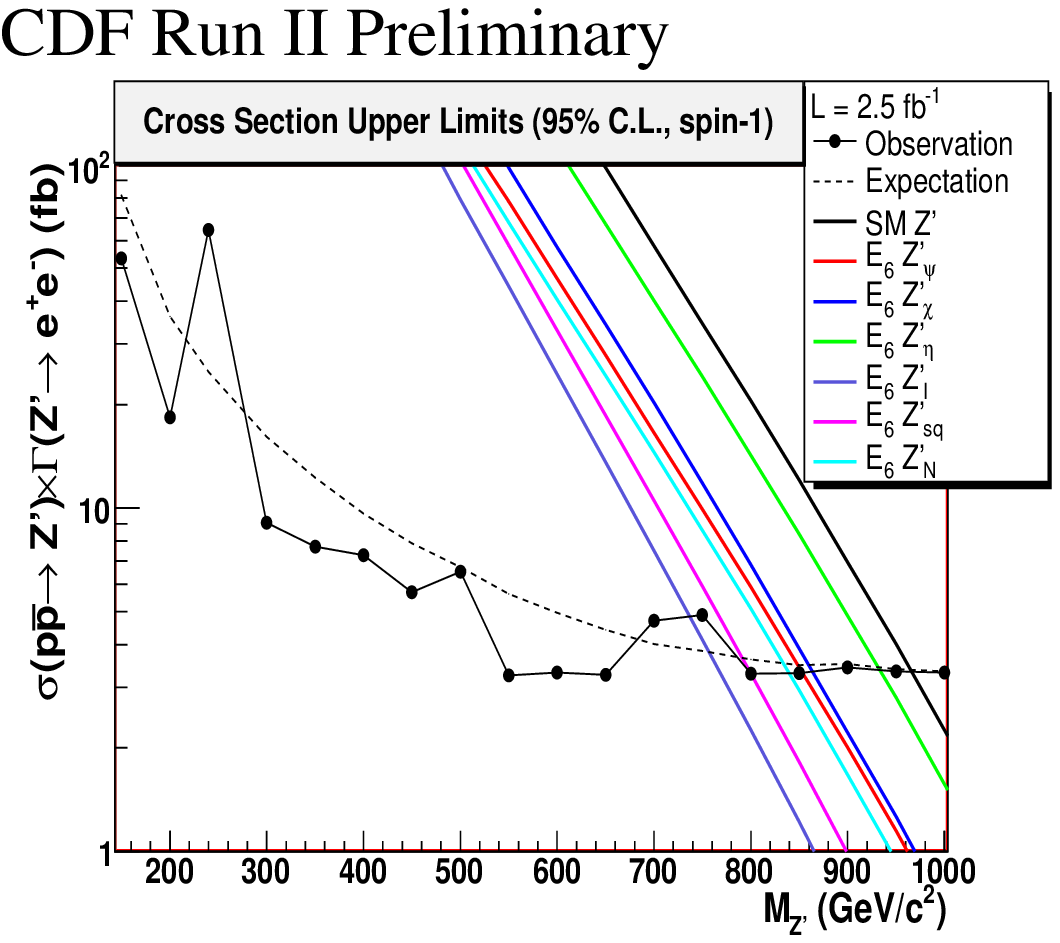,height=2.6in}
\end{minipage}
\begin{minipage}{0.5\textwidth}
  \centering
  \psfig{figure=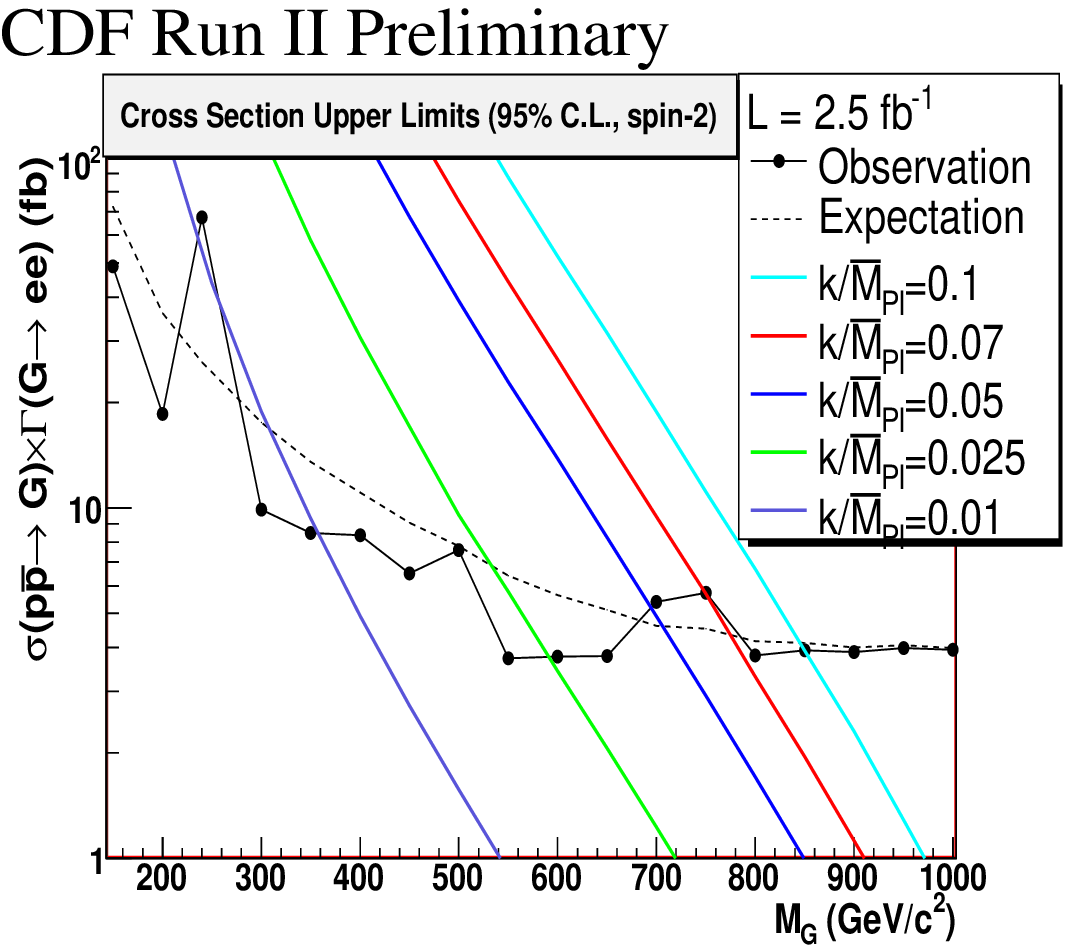,height=2.6in}
\end{minipage}
\caption{95\% CL limit from the CDF dielectron resonance search for
various $Z'$ bosons (left) and RS gravitons (right).
\label{fig:RS}}
\end{figure}

Jet or photon plus missing transverse momentum signatures have been used to 
search for large extra dimensions; the new limits on the scale 
$M_D$ now range from 
1420/1160/1060/990/950 GeV for 2/3/4/5/6 extra dimensions, according to the 
CDF measurements. For RS gravitons the range 850 (350) GeV is excluded
for $k/M_{Pl} = 0.1 (0.01)$, see Fig.~\ref{fig:RS}. New Gauge bosons 
a la Z' are excluded in the range 
below 750 GeV to 1 TeV, depending on the model.

\begin{figure}[htb]
\begin{center}
\includegraphics[angle=-90,width=4.0in]{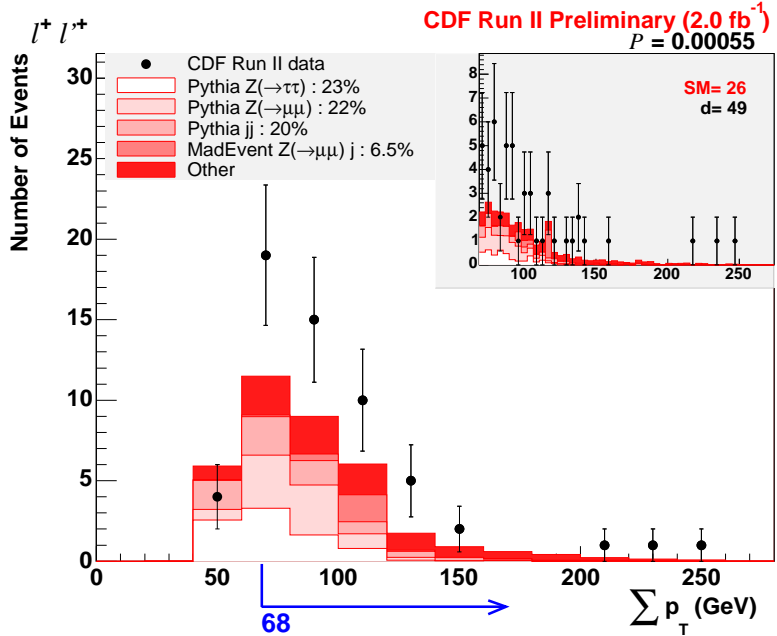}
\caption{Sum $p_T$ for like sing leptons in the CDF data for 2.0 
fb$^{-1}$
as found by \Sleuth\ in 
The region with the most
 significant excess of data over SM expectation is indicated by the blue line
and displayed in the inset. The significance of the excess is shown by $P$.}
\label{fig:sleuth}
\end{center}
\end{figure}

With the advent of the LHC and the plethora of possible new physics 
scenarios, one can wonder if it is possible that we will miss a prominent
signal simply because we didn't think if looking in a specific, perhaps 
weird, channel. Do we need automatic tools for "discovering new physics"?
Several attempt have been made in that direction since a number of years,
with so called generic searches. At this meeting a detailed exposure
on a package of tools for tackling new, basically unknown,  data
was reported`~\cite{henderson}, namely the VISTA package, complemented with 
SLEUTH and Bump Hunter. The tool has been used recently on CDF data and 
after  considerable effort to understand all features in data (including
non-collision background etc), applied to search for new 
physics. At the end a number of discrepancies 
with the data --not related to e.g. insufficient QCD modeling-- have been
identified. An example is shown in Fig.~\ref{fig:sleuth} showing the 
summed transverse momentum of like sign leptons, clearly overshooting 
the data. So far no discovery has been claimed for this excess, however ...

Once the discoveries are made, it will be important to disentangle the 
signatures and map these to theory space to extract the underlying theory.
This is sometimes also called the inverse problem.
In the last few years several attempts are made to test this mapping
~\cite{nima}, look for footprints\cite{kane}, set up 
dictionaries~\cite{belaev}, providing tools to tackle 
such questions from data~\cite{schuster}  and more. 
I believe such exercises have been useful and such tools will
be  needed once new signatures, less trivial than e.g. a Z', will show up 
in the early LHC running.

\section{Finally: The LHC}
The LHC is probably the most complex and challenging scientific 
instrument ever made by mankind. After a long wait, it finally will turn into
operation in 2008, and its start-up is highly anticipated by the particle
physics community. Next year's Moriond meeting should contain LHC data!

It is unlikely --but not entirely excluded-- that the data of 2008
will reveal exciting discoveries, more so since this year we expect that the
top energy of the machine will be 10 TeV instead of 14 TeV, due to some 
magnets that will need retraining during shutdown after the pilot run.
The expected luminosity delivered to the experiments is about 
40 pb$^{-1}$ for 2008, with a large margin of uncertainty of course.

At this conference, many presentations were made on the expectations with 
first data of the LHC and on strategies for 
searches~\cite{Cardaci,yma,rey,black,coco,ambrog,walk,blum,ior,masro,dockt}.
These have been 
presented at many conferences in the past, but in recent years the
attention of the experiments has turned to more data driven techniques 
for estimating backgrounds and efficiencies, and full simulation of the 
different channels.

\begin{figure}[!Hhtb]
\begin{center}\includegraphics[width=0.70\columnwidth]{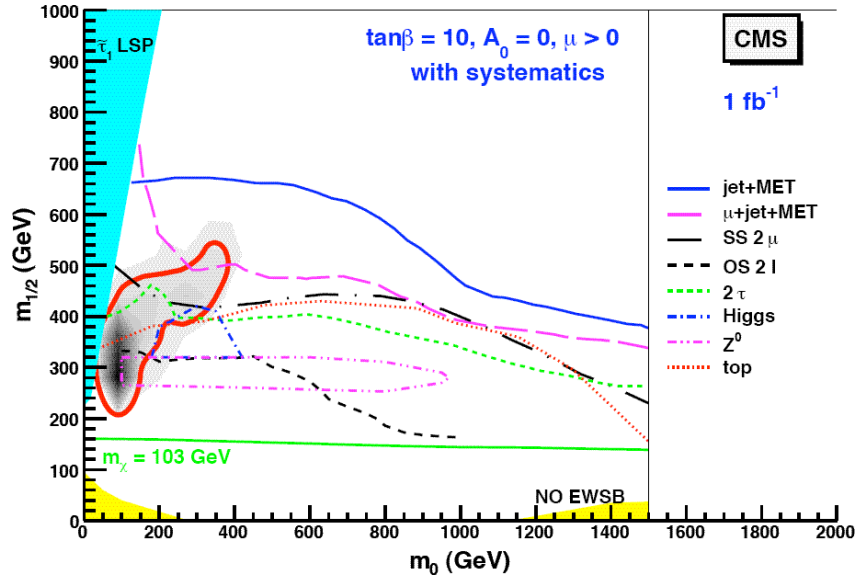}
\end{center}
  \caption{\label{fig:susyreach}
Regions of the $m_0-m_{1/2}$ plane showing the CMS reach with 1 fb$^{-1}$.
The dark region represents the most favoured fit to precision data (see  
text).}
\end{figure}

Early discoveries are possible at the LHC; take e.g. supersymmetry.
The reach in SUSY parameter  space that can be covered by the early 
measurements is
typically studied for benchmark  scenarios. Fig.~\ref{fig:susyreach} shows
that reach for
different final state signatures,  as function of two mSUGRA model
parameters, namely the Universal
scalar and gaugino masses: $m_0$ and $m_{1/2}$.
 The early  reach of the LHC will be large, as already anticipated from
the cross sections given above.
The dark region at low $m_0$ shows the "preferred" region based on a fit
of present precision data and heavy flavour variables within the
constrained MSSM~\cite{buchmuller}. Clearly this region will be probed
already with the first data.

\begin{figure}[htb]
\begin{center}
\includegraphics[width=4.0in]{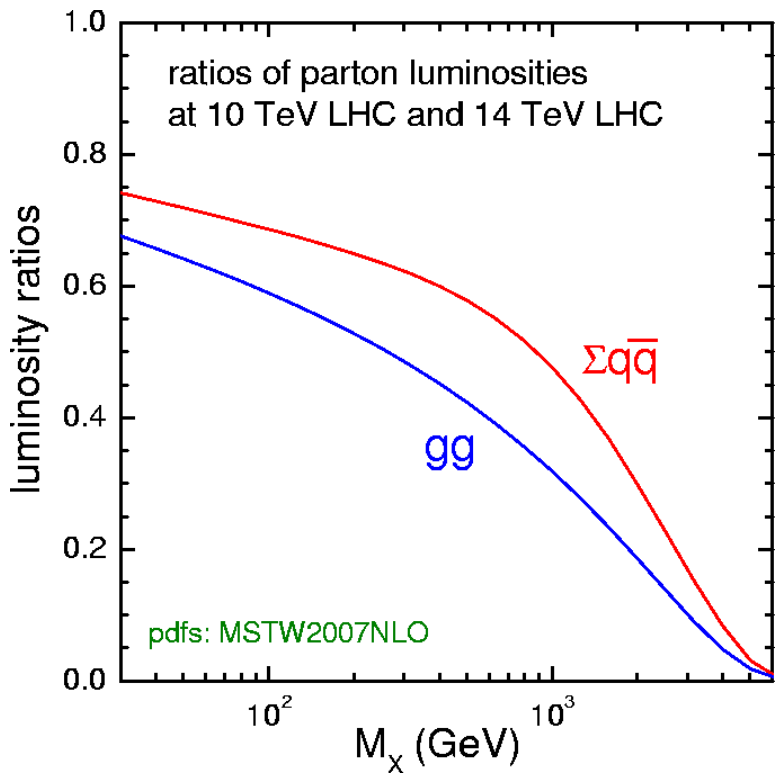}
\caption{The ratio of the cross sections for 10 TeV to 14 TeV for quark quark
and gluon gluon processes in pp collisions at the LHC, from 
\protect\cite{stirling}}
\label{fig:10tev}
\end{center}
\end{figure}

As it got announced that the startup energy of the machine will be 10 TeV
the prospects of these predictions will change. The global effect can be 
anticipated from Fig.~\ref{fig:10tev}, which shows the ratio of the 
cross sections for 10 TeV to 14 TeV for quark-quark and gluon-gluon processes.
In the area for discoveries, say above a TeV, the cross sections 
typically go down by a factor two or more.

Let me end by giving one example of additions that are proposed already now
to the baseline detectors. My completely unbiased choice fell on the 
FP420 project as discussed in~\cite{rouby}. This project proposes and 
extension of the ATLAS and/or CMS baseline detectors by 
putting detectors at 420m away from the interaction point
for protons that have lost less that 1\% of their energy in the interaction
but otherwise remain intact. A full R\&D report on how to do this in
practice is now available\cite{fp420}. From the physics side it will
not only allow CMS and ATLAS to make a number of uncanny QCD, two-photon and 
diffractive measurements due to the extra coverage, but will 
possibly also open a window to study 
properties of the Higgs, such as spin quantum numbers
or --thanks to selections rules-- the  $b\overline{b}$
decay mode and coupling,
 otherwise difficult or impossible 
to access with the 
baseline LHC detectors~\cite{fp420,khoze}. 
The key process here is $pp\rightarrow p+H+p$, i.e. exclusive central 
Higgs production.

\section{Conclusions}
It has been a very lively Moriond QCD 2008, with lots of good data and 
discussions to 
remember, including a first showing of the $F_L$ from HERA, 
the new top mass
 determination and corresponding EW fit results, and
a new Higgs search limit, starting to scratch the area of sensitivity
to the SM Higgs. Some signatures of BSM physics, a bit larger than 
3$\sigma$, have surfaced but we have to see if these is will survive  
further scrutiny and more data. 

But one thing is clear: the LHC is coming this fall!
Hence Moriond 2009 promises to be yet again a very interesting 
meeting.


\section*{Acknowledgments}
I would like to thank the organizers for the invitation and for 
organizing at times real bad 
weather, so I did not feel alone, not skiing. I do wish to congratulate 
them for organizing a perfect conference.

\end{document}